# E-SPEED GOVERNORS FOR PUBLIC TRANSPORT VEHICLES

**[1]C.S.Sridhar,**   **[2]Dr. R.ShashiKumar,**   **[3]Dr.S.Madhava Kumar,**          **[4]Manjula Sridhar,**          **[5]Varun.D**

Asst.prof          Professor          Professor                    Patent-engg          (MSc-Engg)

ECE dept,   SJCIT, Chikkaballapur.                              Teles-AG

**Abstract: -** An accident is unexpected, unusual, unintended and identifiable external event which occurs at any place and at any time. The major concern faced by the government and traffic officials is over speeding at limited speed zones like hospitals, schools or residential places leading to causalities and more deaths on the roads. Hence the speed of the vehicles is to be regulated and confined to the limits as prescribed by the traffic regulations. In this paper we propose a solution in the form of providing E-speed governor fitted with a wireless communication system consisting of a Rx which receives the information regarding the speed regulation for their zones. The TX will be made highly intelligent and decide when receiver should be made active to regulate the speed and unwarranted honking from the vehicles which can be deactivated in the silent zones.

## 1. INTRODUCTION

A wireless transmitter is placed in a particular zone like hospital, school, office or residence. These transmitters are intern monitored by a Personal Computer (PC) which checks the parameters such as time, emergency case and traffic congestion. The PC sends a signal to the transmitter module to set a definite speed depending on these parameters. In case of hospital zone, when Transmitter module detects an emergency signal, it sets speed limit to a predefined value. Similarly for other zones like school, office, residential area ete. The receiver receives the signal and regulates the speed of the vehicle depending on the speed limit of the zone. After the vehicle passes out of the zone the vehicle returns to its normal speed. The gear mechanism facilitates the user to shift gears at lower speeds and also a mechanism is provided to jam all horns in the zone so as to reduce public nuisance and also noise pollution as honking has become a habit in all metros. In case of any chance of collision, the sensor detects the obstruction and halts the vehicle preventing dents on the vehicles. This work is more reliable and robust when compared to an existing mechanical speed governor. This E-SPEED GOVENORS can be used as a total solution for traffic jams, controlling over speeding, providing noise free zone and also in avoiding accidents. Speed governor comprises of two major blocks such as transmitter module and receiver module as shown in figure 1.and is explained in the next section.

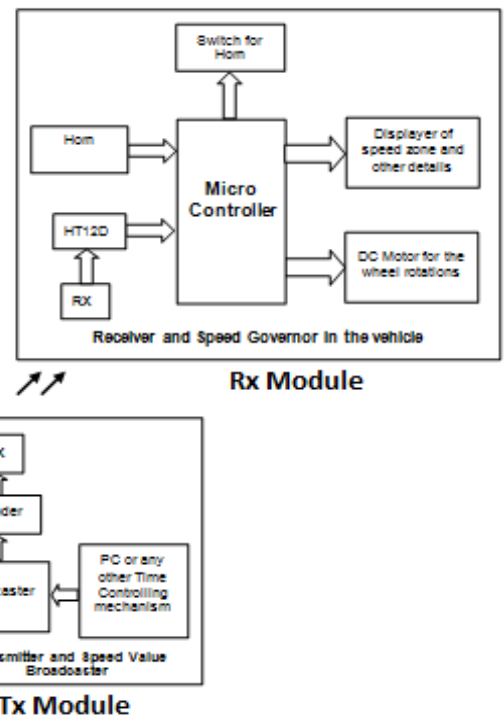

**Figure 1: Block diagram of E-Speed Governor**

## 2. TRANSMITER MODULE

The transmitter defines the speed of the vehicle for the specific zones like school, office or the hospital and sends a special symbol for a speed limit defined for that particular zone. The front end Graphic User Interface (GUI) i.e. user graphic interface is designed using visual basic (VB). In GUI, we can open either COM1 or COM2 for communication. Three zones are defined in the GUI; one can set easily closing and opening time for each zone. The interface between the PC and microcontroller is done by RS- 232 serial port. The symbol sent by the PC is received by the microcontroller for further processing.

The transmitter transmits special symbols such as "!","@","#","$","%","^" which helps in differentiating zones and corresponding zones speed. The symbol is sent through the RS 232 port to the microcontroller in which it is processed and encoded in an encoder and transmitted through a transmission line pack (TLP) in different zones. The signal is transmitted at 433.93 MHz after suitable modulation and

amplification. The transmitter modules with its subsystem are explained in the following section.

## 2.1 RF transmitter Module [4]

The RF transmitter uses the frequency modulation technique for transmitting. the signals to the receiver circuit or the governor which is placed inside the vehicle. The frequency range falls in VHF [Very High Frequency] and hence resembles the commercial FM broadcasting system. These signals are then aired by using the antenna coil.

## 2.2 Power supply

Power required to drive the PC and the microcontroller is provided by two distinct SMPS on the cabinet and on the circuit board [car battery in case of real time example].

## 2.3 Control Unit

The block diagram of control unit is as shown in figure 2.It gives the over-all description about how the received signal [to the microcontroller] is encoded and applied on the vehicle whose speed is to be governed. The controlling part can be sub-divided into following sections: data bus, Transmitter microcontroller, encoder and transmitter line pack.

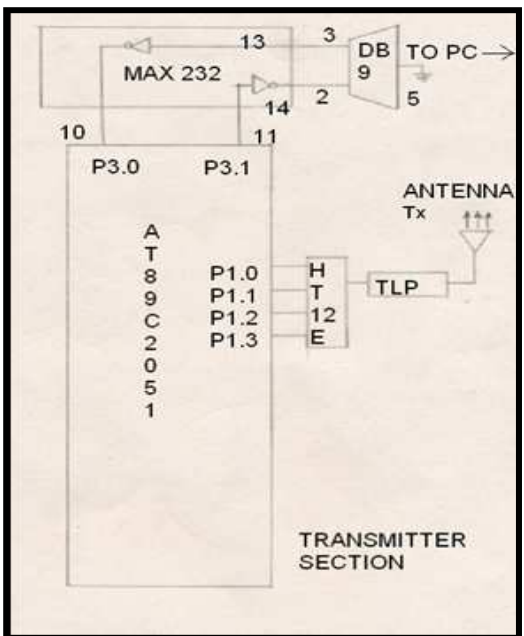

**Figure.2 Control Unit**

### 2.3.1 Data BUS [RS232]

The Serial Port RS 232 is used as interface between PC and transmitter module. Both synchronous and asynchronous transmissions are supported by this standard. Any signal transmitted by the PC is received by the microcontroller through this RS-232 serial port. The voltage levels are made higher than logic levels using MAX-232.

### 2.3.2 Transmitter Microcontroller [5]

In the transmitter section, we require a microcontroller to analyze the transmitted data from the PC. The low cost AT89C2051 microcontroller is used for this purpose. The PC selects the zone and sends signal to the Microcontroller through RS232 port. The received signal is analyzed and fed to encoding block by the microcontroller..

The AT89C2051 is a low-voltage, high-performance CMOS 8-bit microcomputer with 2K bytes of Flash EPROM.

### 2.3.3 Encoder [HT12E]

The HT12E encoders are a series of CMOS LSIs for remote control system applications. They are capable of encoding information which consists of 8 address bits and 4 data bits. Each address or data input can be set to one of the two logic states. The programmed address or data are transmitted together with the header bits via an RF or an infrared transmission medium upon receipt of a trigger signal. The pin details of HT12E and connection with RF transmitter is as shown in figure.3

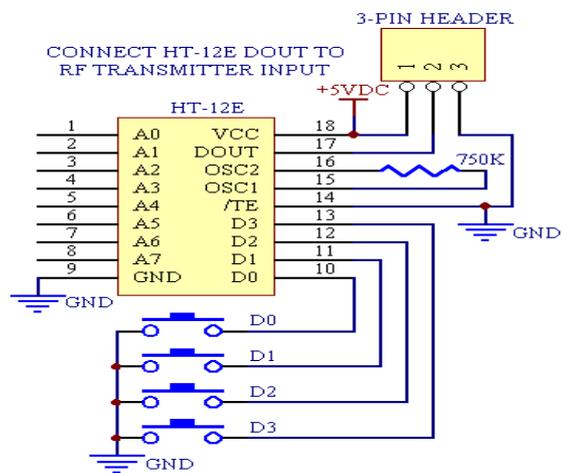

**Figure.3 Encoder Unit**

### 2.3.4 Transmitter Line Pack [TLP434A] [4]

TLP 434A is used to transmit a frequency of 433.93 MHz. This is required to transmit the message to control the speed and the horn of the vehicle. Figure 4 shows the interface of TLP434A with HT12E

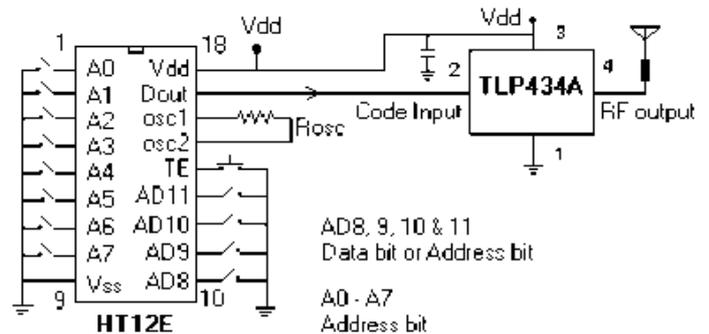

**Figure.4 Interface of TLP434A with HT12E**

## 3   RECEIVER MODULE [7]

The antenna receives the modulated signals of each zone and is demodulated by RLP (receiver line pack) which demodulates the received signal and pre amplifies it and sends to the decoder [HT12D] to decode signal. The decoded information is fed to the microcontroller unit and depending upon the received information the microcontroller decides the speed of the vehicle. LCD enables us to know the speed of the vehicle and also the zone in which the vehicle is traveling.  This in turn reduces the speed of the dc motor which we have used for the demo purpose. We have a gear mechanism to switch between speed and torque. A horn jamming system is fixed which jams the horn in non honking zone. Accident avoidance is also provided for the safety of the vehicle and pedestrian.  Block diagram of the receiver is shown in figure.5

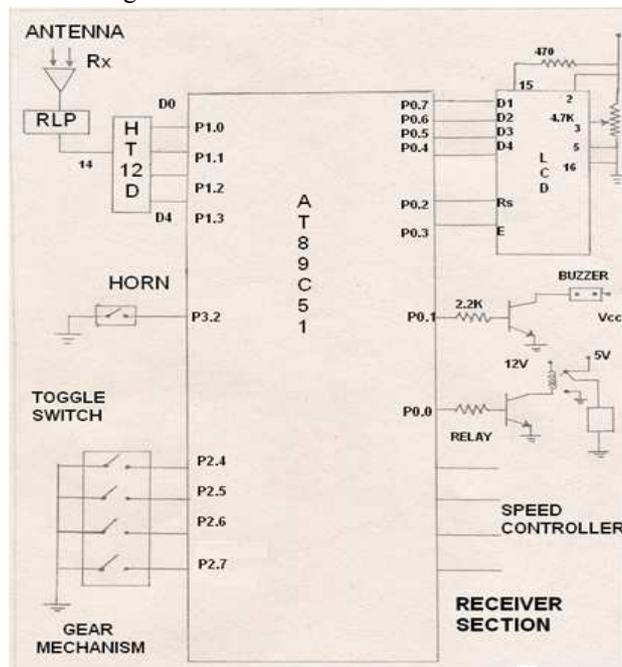

**Figure.5 Receiver module**

### 3.1  Receiver Line Pack [9]

A receiver is required to receive the speed governing message sent by Transmitter at the desired zones. Hence the receiver must receive signal of the same frequency as that of the transmitter. The receiver then sends the information to the decoder which decodes and sends the signal to the microcontroller for further processing.

### 3.2  Decoder [HT12D] [6]

A 4 bit (HT 12D)is used to decode the serially received signal which is in the analog form. The 2^12 series of decoders provides various combinations of addresses and data pins in different packages so as to pair with the 2^12 series of encoders. The decoders receive data that are transmitted by an encoder and interpret the first N bits of code period as addresses and the last 12_N bits as data, where N is the address code number. A signal on the DIN pin activates the oscillator which in turn decodes the incoming address and data. The decoders will then check the received address three times continuously. If the received address codes match all the contents of the decoders, local address, the 12_N bits of data are decoded to activate the output pins and the VT pin is set high to indicate a valid transmission.

### 3.3  Receiver Microcontroller [1,2]

A microcontroller takes full responsibility of controlling the characteristics of the vehicle. Microcontroller AT89C51 is used to facilitate the control of the vehicle i.e. control of speed, horn-jamming, accident avoidance, display and data processing.

### 3.4  Gear Mechanism

In an internal combustion 4-stroke engine there needs to be a gear box to control the speed and torque of the vehicle depending on terrain. Here a gear mechanism is provided using toggle switches which is directly connected to the microcontroller which would intern trigger relays which would increase the speed. The gear mechanism can also be replaced by SCR triggering which would trigger depending on the amount of acceleration given through a set of toggle switches.

### 3.5  Motor

A prototype motor is used to demonstrate the effectiveness in speed governance and also accident avoidance. Essentially a DC motor is used for practical purposes [as in case of a real time example an internal combustion engine is used].  It consists of a voltage divider circuit which is required to apply different voltages at different instants for different speeds. The voltage is suitably altered for the relay to turn on which is governed to show following parameters

- Sensor placed for accident avoidance.
- Gear mechanism.
- Microcontroller which receives the zone regulated  speeds.

The relay circuit is essentially a switching circuit which is required to switch on the motor. ULN2003 IC used as relay driver.

### 3.6  Horn Control

Horn is an integral part of a vehicle and is used in the vehicle to alert the drivers or pedestrians in front to move aside and leave some space. However the use of horn has become a nuisance and people don't understand it and still unnecessarily honk and which in turn causes noise pollution and other hearing loss. A jamming circuit is established to jams the horn in certain honk free zones.

A horn circuit is connected to the microcontroller.. The switching is provided by a push button which would send logic 1 to the microcontroller [at one of its ports]. When the switch is pressed this causes the microcontroller to send logic 0 [low] at the base of transistor (2N3904) [base connected to one of the ports] this would intern switch on the buzzer. When in a zone the ports are masked and hence would result in jamming of horn.

## 4 IMPLEMENTATION AND RESULT [3,8,9]

The complete code is developed by using Assembly language and Keil micro-vision cross compiler was used for code compilation. The code is optimized using less than 2K memory and the code is well structured and easy for further implementation and modifications. The front end Graphic user interface (GUI) is developed by using Visual Basic (VB) software. The parameters can be varied to meet the changing traffic regulations by incorporating changes in the software.

Figure 5 shows the snapshots of results implemented in real time at office zone during peak hours with opening and closing time specified and with a speed regulation of 45 Kmph.

**Figure.6 Office zone**

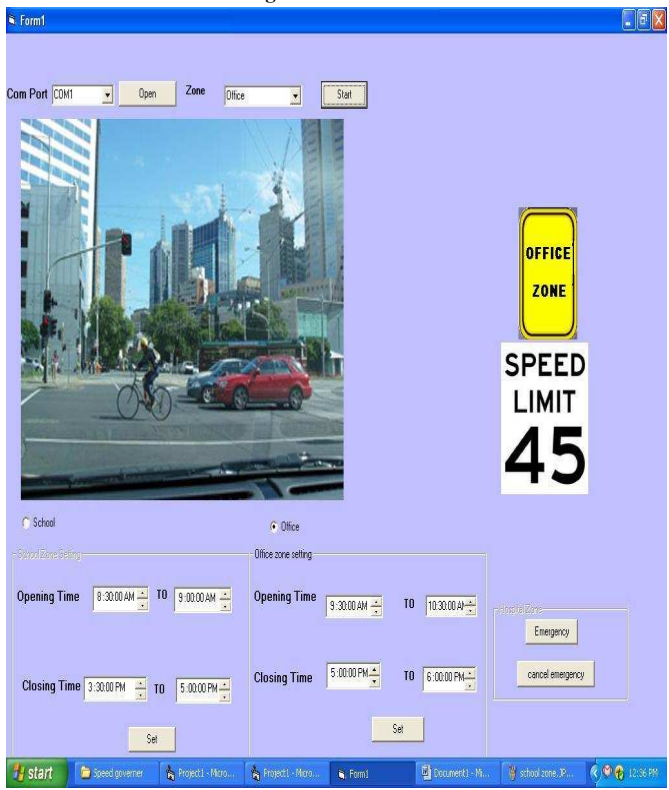

Figure 6 shows snapshot of speed governors implemented at hospital zone with round the clock speed regulation and with the option of deactivation during the times of emergency.

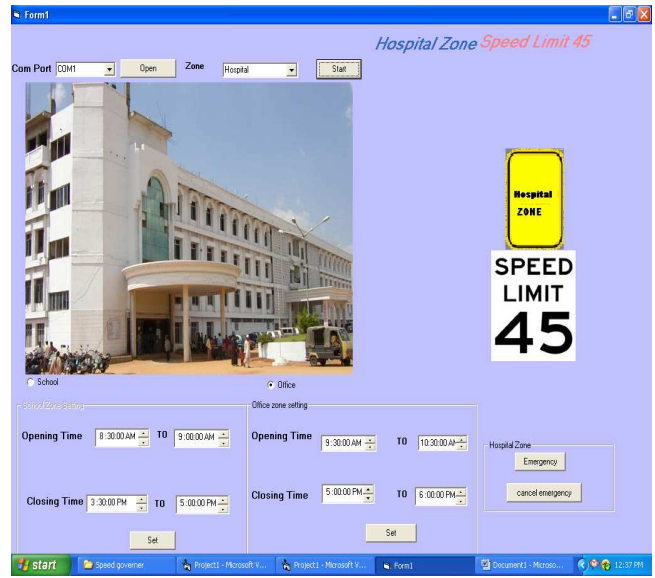

**Figure.7 Hospital zone**

Figure 7 shows the snapshot of speed governors implemented at school zone during the opening and closing hours of school.

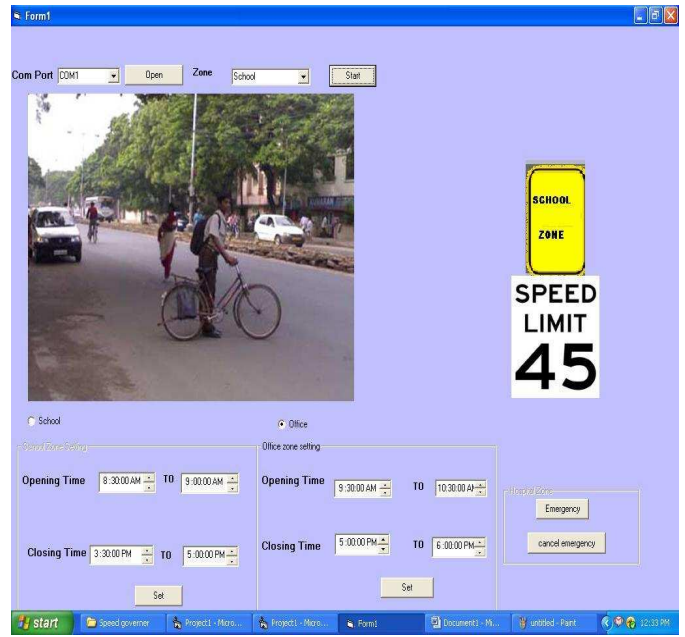

**Figure.8 School zone**

Speed governor was implemented in real time at school zone during opening and closing hours of schools specified with a speed regulation of 45KmpH.

## 5 CONCLUSION AND REMARKS

This work is more reliable and effective in curbing over speeding, accident avoidance and also to provide noise free zones. The implementation and installation is easy and almost needs no maintenance. In case of co-located zones there may be chances of co- channel interference which can be avoided by allocating different frequencies. In case of snow fall there may be chances of vehicles colliding each other due to slippery conditions. All vehicles which are fitted with receivers need to have a CAN controller for auto transmission purposes. Frequency allocation comes under FCC regulation.

RFID/GSM technology can be implemented in place of a transmitter and receiver. Accident avoidance can be done by implementing a transmitter of different frequency and a corresponding receiver in the other vehicle. Back light at the time of entering the zone irrespective of the driver applying brakes or changing gears can be done to avert accidents. LCD boards can be provided at each zone specifying the speed limit and also the zone length can be specified.

## AUTHORS PROFILE

[1] Sridhar .C.S is an engineer from Electronics and

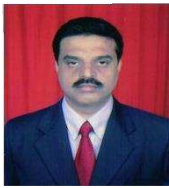

Communications. He has done Masters Degree in Communication Networking and security. Presently he is working as Assistant - Professor in the department of ECE, SJCIT, Chickballapur. His areas of interest are embedded system, wireless and Networking.

[2] Dr. R. Shashikumar is presently working as a Professor

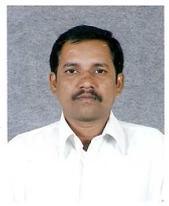

in E & C dept, SJCIT, Chikballapur, Karnataka, India. He is having 10 years of teaching and 6 years of Industry experience. His areas of interest includes ASIC, FPGA, Network Security.

[3] Dr.S.Madhavakumar graduated in electrical engineer from

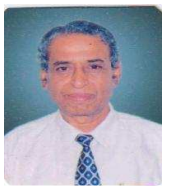

university of Mysore in 1964 and obtained his post graduation and doctoral degree from the University of Roorkee in 1973 and 1984 respectively specializing in communication systems. He has worked as a faculty member in

PES College of engineering, Mandya in various capacities from 1965 to 1987 and worked in several institutions as professor and Head of ECE department. He has also worked as Dean, Sikkiim manipal institute of tech, Sikkim during 2000-08 and presently working as HOD of ECE department, SJCIT, Chikballapur. He is a fellow of IETE and a life member of ISTE and his field of interests are signal processing and communications.

[4] Mrs. Manjula Sridhar has completed MSc in I.T. She is

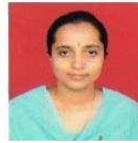

presently working as patent-engineer for Teles-AG, India. Her fields of interest are Java, IPR etc.

[5] VARUN .D has done his BE in Telecommunication and

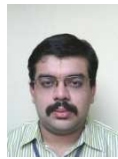

presently perusing his M.Sc [Engg] in Signal Processing & Communication Technologies. His areas of interest are Microwave Communication, Signal Processing & Wireless Communication